\documentclass{PoS}

\usepackage{booktabs}

\newcommand{\ve}[1]{\ensuremath{\mathbf{#1}}}
\newcommand{\n}[1]{\ensuremath{|\mathbf{#1}|}}

\graphicspath{{plot/}}

\title{Breakdown of the impulse approximation\\ and its consequences: the low-$Q^2$ problem}

\ShortTitle{Breakdown of the impulse approximation}

\author{Artur M. Ankowski\\
        Institute of Theoretical Physics\\ University of Wroc{\l}aw\\ pl. Maksa Borna 9\\ 50-204 Wroc{\l}aw\\ Poland\\
        E-mail: \email{artank@ift.uni.wroc.pl}}

\abstract{Neutrino scattering data and the standard calculations of the cross section show a discrepancy in the low-$Q^2$ (four-momentum transfer squared) region. The calculations rely on the assumption, called the impulse approximation, that the nucleus the neutrino scatters off can be described as a~collection of independent nucleons and therefore only one nucleon takes part in the interaction. It is known from electron scattering that such picture is valid only when transferred momentum $\n q\gtrsim400~\textrm{MeV}/c$. For lower $\n q$'s, the nucleus is probed with a~lower spatial resolution and a few nucleons are involved in the scattering, so the use of the impulse approximation is unjustified. It means that the standard calculations of the cross sections are unreliable for $\n q\lesssim400~\textrm{MeV}/c$. I show that the contribution of low-momentum interactions to the quasielastic $\nu_\mu$ cross section cannot be reduced below $\sim$19\% by any selection of the beam energy in the few-GeV region. Moreover, low-$Q^2$ neutrino events happen only when transferred momentum is low too. Hence the discrepancy in the low-$Q^2$ region is caused by the inappropriate description of low-$\n q$ interactions in the models commonly used in Monte Carlo simulations.}

\FullConference{10th International Workshop on Neutrino Factories, Super beams and Beta beams\\
		 June 30 - July 5 2008\\
		 Valencia, Spain}

\begin{document}

When lepton transfers momentum $\ve q$ to the nucleus it interacts with, a~region of the nucleus of order $\sim$$1/\n q$ is penetrated. For low values of $\n q$, the region covers more than one nucleon and, inevitably, a~few particles are involved in the scattering. For higher momentum, the region is small enough to treat the nucleus as a~set of independent nucleons. It is then justified to describe the interaction as a~scattering off a~single (bound) nucleon. Such approach is called the impulse approximation (IA). Within the IA scheme, the nucleus is fully characterized by its spectral function, i.e. the function which describes the distribution of the momenta and energies of the nucleons that compose the nucleus. From the experience gained in electron scattering one knows that the IA works well for $\n q\gtrsim400~\textrm{MeV}/c$ (as far as realistic spectral functions are concerned) and obviously fails below 300 MeV$/c$~\cite{ref:Ankowski&Sobczyk}. Since a~special case of the spectral function is the Fermi gas model, the IA is a~standard approach in which one calculates the neutrino-nucleus cross section.

The results presented in this paper concern the oxygen nucleus described by the spectral function of Ref.~\cite{ref:Benhar&Fabrocini&Fantoni&Sick}. The conclusions, however, hold generally for the IA scheme. I consider quasielastic scattering of muon neutrinos only and neglect the effect of final state interactions.



In neutrino physics, due to poor statistics, important observables are the total cross section and the differential cross section in four-momentum transfer squared $Q^2=\ve q^2-\omega^2$. It is important to know how these quantities calculated in the IA scheme are influenced by the inadequate description of low-$\n q$ events.

\begin{figure}
    \centering
    \includegraphics[width=0.53\textwidth]{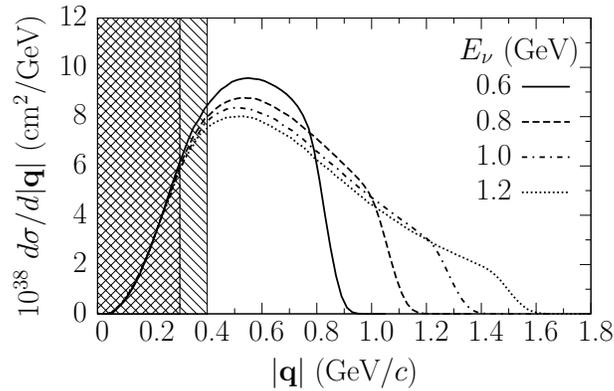}%
\caption{\label{fig:nuOxq} Differential cross section $d\sigma/d\n q$ of the process $^{16}$O$(\nu_\mu,\mu^-)$ at various neutrino energies.}
\end{figure}

When $\n q<300~\textrm{MeV}/c$, the differential cross section $d\sigma/d\n q$ is almost identical for neutrino energies $E_\nu\geq600$~MeV, see Fig.~\ref{fig:nuOxq}. Therefore the contribution of low $\n q$'s to the total cross section is constant and significant in the whole few-GeV region, see Tables~\ref{tab:IABreakdown} and~\ref{tab:IABreakdown2}, and reaches its minimum at $E_\nu\sim1550$~MeV, i.e. at the energy corresponding to the maximal cross section.

\begin{figure}
    \centering\vspace{1em}
    \includegraphics[width=0.53\textwidth]{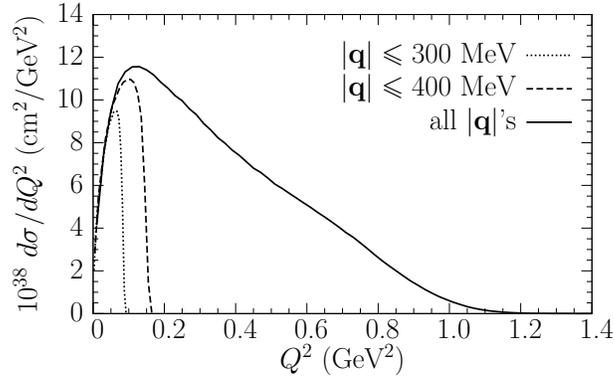}%
\caption{\label{fig:nuOxQ2} Contribution of low momentum transfers to the differential cross section $d\sigma/d Q^2$ of the $^{16}$O$(\nu_\mu,\mu^-)$ scattering at neutrino energy 800 MeV.}
\end{figure}

Comparison of theoretical results obtained within the IA to experimental data exhibits a~discrepancy in the low-$Q^2$ region, namely the calculations overestimate event yields~\cite{ref:K2K,ref:MiniB}. As shown in Fig.~\ref{fig:nuOxQ2}, whole the cross section at low $Q^2$ comes from low momentum transfers, i.e. from the region where one cannot rely on the IA. Such clear situation allows identification of the cause of the discrepancy as the breakdown of the IA. A~correct description of low-$\n q$ interactions, e.g. within the random phase approximation, should reduce the cross section in the low-$Q^2$ region.

One should keep in mind that the breakdown of the IA introduces additional uncertainty to the models commonly used in Monte Carlo simulations.

\begin{table}
\vspace{0.5em}
\centering
    \begin{tabular}{@{}lrrrrrrr@{}}
    \toprule
    & \multicolumn{7}{c}{Neutrino energy (GeV)}\\
    \cmidrule(l){2-8}
    &\multicolumn{1}{c}{0.2} & \multicolumn{1}{c}{0.4} & \multicolumn{1}{c}{0.6} & \multicolumn{1}{c}{0.8} & \multicolumn{1}{c}{1.0} & \multicolumn{1}{c}{1.2}& \multicolumn{1}{c}{1.4}\\
    \cmidrule(l){2-8}
    $\n q\leq300$ MeV$/c$  &  97.2\% & 18.9\% & 11.9\% & 10.1\% & 9.4\% & 9.1\% & 9.0\%\\
    $\n q\leq400$ MeV$/c$  & 100.0\% & 43.3\% & 26.2\% & 21.6\% & 19.8\%& 19.1\%&18.8\%\\
    \bottomrule
    \end{tabular}
    \caption{Low-$\n q$ contribution to the neutrino cross section.}
    \label{tab:IABreakdown}
\end{table}

\begin{table}
\vspace{0.5em}
\centering
    \begin{tabular}{@{}lrrrrrrr@{}}
    \toprule
    & \multicolumn{7}{c}{Neutrino energy (GeV)}\\
    \cmidrule(l){2-8}
    &\multicolumn{1}{c}{2.0} & \multicolumn{1}{c}{2.5} & \multicolumn{1}{c}{3.0} & \multicolumn{1}{c}{3.5} & \multicolumn{1}{c}{4.0} & \multicolumn{1}{c}{4.5}& \multicolumn{1}{c}{5.0}\\
    \cmidrule(l){2-8}
    $\n q\leq300$ MeV$/c$  &  9.1\% & 9.2\% & 9.3\% & 9.3\% & 9.4\% & 9.4\% & 9.5\%\\
    $\n q\leq400$ MeV$/c$  & 18.8\% & 18.9\% & 19.1\% & 19.1\% & 19.3\%& 19.3\%&19.4\%\\
    \bottomrule
    \end{tabular}
    \caption{Same as Table~1 but for higher neutrino energies.}
    \label{tab:IABreakdown2}
\end{table}

\acknowledgments
I am deeply indebt to Jan T. Sobczyk for discussions on the subject of this paper and to Omar Benhar for the spectral function of oxygen. This work was supported by MNiSW under grant nos. 3735/H03/2006/31 and 3951/B/H03/2007/33.

\end{document}